\documentclass[a4paper]{PoS}

\usepackage{epsf}

\def\NPB#1#2#3{{\it Nucl.\ Phys.}\/ {\bf B#1} (#2) #3}
\def\PLB#1#2#3{{\it Phys.\ Lett.}\/ {\bf B#1} (#2) #3}

\def\PRD#1#2#3{{\it Phys.\ Rev.}\/ {\bf D#1}  (#2) #3}
\def\PRL#1#2#3{{\it Phys.\ Rev.\ Lett.}\/ {\bf #1} (#2) #3}
\def\PRT#1#2#3{ {\it Phys.\ Rep.}\/ {\bf#1} (#2) #3}
\def\MODA#1#2#3{ {\it Mod.\ Phys.\ Lett.}\/ {\bf A#1} (#2) #3}

\def\IJMP#1#2#3{ {\it Int.\ J.\ Mod.\ Phys.}\/ {\bf A#1} (#2) #3}

\def\EJP#1#2#3{ {\it Eur.\ Phys.\ Jour.}\/ {\bf C#1} (#2) #3}
\def\JHP#1#2#3{ {\it JHEP}\/ {\bf #1} (#2) #3}

\def\etal{{\it et al\/}}

\def\AEF{A.E. Faraggi}

\newcommand{\nn}{\nonumber}
\newcommand{\beq}{\begin{equation}}
\newcommand{\eeq}{\end{equation}}
\newcommand{\beqa}{\begin{eqnarray}}
\newcommand{\beqn}{\begin{eqnarray}}
\newcommand{\eeqn}{\end{eqnarray}}
\newcommand{\eeqa}{\end{eqnarray}}
\newcommand{\oo}[2]{\left(#1\left|#2\right.\right)}
\newcommand{\ba}{\begin{eqnarray}}
\newcommand{\ea}{\end{eqnarray}}

\title{Spinor-vector duality and light Z' in heterotic strings}

\ShortTitle{Spinor-vector duality and light Z' in heterotic string vacua}

\author{\speaker{Alon E. Faraggi}\\%
        Mathematical Sciences Department, 
        University of Liverpool, Liverpool L69 7ZL, UK\\
        E-mail: \email{Alon.Faraggi@liverpool.ac.uk}}

\author{John Rizos\\
       Department of Physics,
              University of Ioannina, GR45110 Ioannina, Greece\\
        E-mail: \email{irizos@uoi.gr}}

\abstract{
We discuss the construction of heterotic--string  models that
allow for the existence of an extra $Z^\prime$ at low scales. 
One of the main difficulties encountered is that the desired symmetries
tend to be anomalous in the prevailing three generation 
constructions. The reason is that these models utilise
the symmetry breaking pattern $E_6\rightarrow SO(10)\times U(1)_\zeta$
by GGSO projections. Consequently, $U(1)_\zeta$ becomes anomalous. 
The spinor--vector duality that was observed in the
fermionic $Z_2\times Z_2$ orbifold compactifications 
is used to construct a phenomenological three generation 
Pati--Salam heterotic--string model in which $U(1)_\zeta$ is anomaly
free and therefore can be a component of a low scale 
$Z^\prime$. The model implies existence of matter states
at the $Z^\prime$ breaking scale, which are required for
anomaly cancelation. Moreover, the string model
gives rise to exotic states, which are $SO(10)$ singlets
but carry exotic $U(1)_\zeta$ charges. These 
states arise due to the breaking 
of $E_6$ by discrete Wilson lines and provide 
natural dark matter candidates. 
Initial indications suggest that the existence of 
additional gauge symmetries at the TeV scale may be confirmed in run II 
of the LHC experiment. 
}

\FullConference{
18th International Conference From the 
Planck Scale to the Electroweak Scale, 
                 25-29 May 2015, 
                 Ioannina, Greece }

\begin{document}

\section{Introduction}

The Standard Model of elementary particle physics continues to 
provide an accurate parameterisation of all
particle physics observations. The data 
from the lepton colliders during the 1990s confirmed the 
non--Abelian nature of the electroweak and strong interactions
to an unprecedented precision. The observation of a Higgs--like
state at the LHC represents another triumph of the model. The task 
of future experiments will be to continue to probe the Standard 
Model parameterisation to better and better precision, and
possibly discover its plausible extensions. It
is also not implausible that the Standard Model is all there
is at the energy scales within reach of the LHC
and possibly within reach of collider experiments
in the decades to come. This does not diminish from the 
vitality of these experiments and their importance. If
nature consists solely of the Standard Model in this 
energy range we ought to confirm that this is the
indeed the case in future experiments. The challenge to
develop the instruments that can probe nature at the increasing
energy scales benefits the societies that pursue this 
endeavour and ultimately offers them technological and 
economic supremacy.
At the present state of affairs the standard model consists 
of three gauge sectors, three generations of matter families
with six chiral multiplets per family\footnote{including
a right--handed neutrino, which is instrumental for the
observed neutrino data}, and a single electroweak
scalar doublet. Perhaps the most striking feature of the 
Standard Model is the fact that its matter content 
fits into chiral 16 representations of $SO(10)$. 
The significance of this coincidence is exemplified most clearly 
if we consider that the Standard Model gauge quantum numbers
are experimental observables. The Standard Model requires fifty--four 
parameters to account for these charges, whereas its embedding
in $SO(10)$ reduces this number to one parameter, being the 
number of $SO(10)$ chiral 16 representations
required to embed the Standard Model states.
The success of the Standard Model provide strong support
for the realisation of the $SO(10)$ unification structures
in nature. The scale where these unification structures 
become relevant is, however, far removed from the 
electroweak scale. 
Evidence to this effect stem from the observed logarithmic running
of the Standard Model parameters; the proton longevity; and 
the suppression of left--handed neutrino masses. 

Despite its enormous success, the Standard Model leaves several gaps. 
While QCD provides an accurate parameterisation 
of the strong interactions in the perturbative regime, a detailed 
understanding of its nonperturbative infrared 
limit and confinement is still lacking.
The gravitational effects are not accounted for. Moreover, there
is a fundamental dichotomy between point quantum field theories, 
the calculational framework underlying the Standard Model, 
and general relativity, which underlie the gravitational interactions. 
String theories produce a self--consistent approach 
to the synthesis of general relativity and quantum mechanics.
Furthermore, the string consistency conditions mandate the existence
of gauge and matter states, similar to those that are present 
in the Standard Model. String theories therefore facilitate
the development of a viable phenomenological approach 
to probe how the gravitational and gauge interactions 
may be reconciled. 


Heterotic string theory is particularly
appealing because it gives rise to spinorial representations
in its perturbative spectrum, and thus enables the embedding 
of the Standard Model matter states in spinorial 
$SO(10)$ representations. The free fermionic formulation \cite{fff} 
of the heterotic string provides a particularly fertile framework
for the construction of phenomenological string vacua
with $SO(10)$ embedding of the Standard Model chiral states. 
It is important to note that the $SO(10)$ symmetry is broken
to one of its subgroups directly at the string scale. This gives rise to 
the string doublet--triplet splitting mechanism. It enables the Higgs
states to exist in incomplete $SO(10)$ multiplets, 
which facilitates the compatibility with the gauge
coupling data at the electroweak scale, as well as with proton 
lifetime limits. The primary guides in the search of quasi--realistic
string vacua are the existence of three chiral generations and their
$SO(10)$ embedding. 

\section{Free fermionic models}

A class of phenomenological string vacua that meet these criteria are 
the quasi--realistic string models in the free fermionic formulation. 
Since the early 1990s this class of models provided a laboratory to study
how the phenomenological features of the Standard Model
may arise from string theory. A few of the highlights include: 
\begin{itemize}
\item Minimal Superstring Standard Model \cite{fny, cfr}. Construction 
of string models leading to solely the MSSM spectrum below the 
string scale. 
\item Top quark mass $\rightarrow 175-180$GeV \cite{top}. Calculation
of the top and bottom quarks Yukawa couplings at the string scale 
yielding a prediction of the top quark mass at the 
electroweak scale. 
\item Fermion masses and CKM mixing \cite{fmmckm}.  
\item Stringy seesaw mechanism and neutrino masses \cite{nmasses}.
\item Gauge coupling unification \cite{gcu}.
\item Proton stability \cite{ps}
\item Squark degeneracy \cite{squarkdegen}.
\item Moduli fixing \cite{moduli}.
\item Classification \& Exophobia \cite{classification, psclassi, fsuclassi}. 
\item Spinor--vector duality \cite{svduality,florakis}.
\end{itemize}
It should be stressed that this free fermionic construction
probes one class heterotic--string vacua, which are related 
to $Z_2\times Z_2$ orbifold compactification. Other classes
of string vacua can be probed by using a variety of tools that 
have been developed over the years. These include geometrical,
orbifolds, interacting world--sheet conformal field theory 
constructions;
orientifolds. A comprehensive review
of different approaches to string phenomenology is
given in ref. \cite{ibanezuranga}. 

\section{$Z^\prime$s in free fermionic models}

One of the well motivated extensions of the Standard Model is 
the existence of an extra gauge symmetry beyond the Standard Model. 
The first argument in favour of such extensions is
that gauge symmetries actually exist in nature, and the
assumption that an additional gauge symmetry exists is not 
outrageous. More concretely, we have argued that the Standard Model
gauge charges strongly hint to the $SO(10)$  embedding 
of the Standard Model matter states. This necessitates that 
the Standard Model gauge symmetry is extended by at least
one $U(1)$ factor. The promotion of the global baryon and lepton
symmetries to a local symmetry further hints that the Standard 
Model gauge symmetry should be extended. This again fits 
well with the $SO(10)$ paradigm in which the
baryon minus lepton number is gauged. Anomaly cancellation in perturbative
string theory, which 
incorporates the Standard Model building blocks and gravity 
in one embrace, mandates the existence of additional gauge symmetries. 
Furthermore, the heterotic--string fuses those ingredients and 
reproduces the $SO(10)$ structure underlying the Standard Model. 
It should be 
emphasised again 
that the $SO(10)$ symmetry is 
not manifested as a gauge symmetry in the effective 
field theory limit of the string vacua, but merely serves
as an organisional framework from that point of view. It's 
ultimate role, once the full string dynamics is better 
understood is a story for the future. Possibly for future 
generations. 

The Standard Model points to the existence of additional 
gauge symmetries, while string theory mandates their 
existence. Alas, the additional gauge symmetries 
may be broken at a high scale compared to the 
electroweak scale and would not be observed in
contemporary collider experiments. Non--Abelian
extensions of the Standard Model like $SU(5)$ 
is consistent with proton decay limits only
if the $SU(5)$ symmetry is broken above $10^{16}GeV$. 
The most plausible extension of the Standard Model
within reach of contemporary collider experiments is 
an Abelian gauge symmetry. 

The heterotic--string models in the free fermionic formulation
reproduce the general features of the Standard Model and provide
a setting to investigate additional gauge symmetries in quasi--realistic 
string constructions.
While additional spacetime vector bosons are abundant in the string 
models, the possibility that they remain massless to low energy
scales is far less obvious. In fact for a variety of reasons most 
of the additional gauge symmetries have to be broken at a scale 
beyond the reach of the LHC. 
Exploration of additional $Z^\prime$s in free fermionic
models started in the early nineties. The first case to be considered
\cite{zpbminusl}
was the combination 
\beq
U(1)_{Z^\prime}={3\over2}U(1)_{B-L}-2 U(1)_R\in SO(10).
\label{u1zpinso10}
\eeq
Existence of this $Z^\prime$ at the TeV scale ensures the 
suppression of proton decay from dimension four operators, 
which are endemic in supersymmetric extensions of the 
Standard Model. However, the underlying $SO(10)$ symmetry 
in the string models dictates that the Dirac mass term of the
tau neutrino is equal that of the top quark. Breaking 
the $U(1)_{Z^\prime}$ symmetry of eq. (\ref{u1zpinso10})
at the TeV scale generates a low scale seesaw, which implies 
either a relatively heavy tau neutrino or that some scalar 
fields get an ad hoc VEV of order $O(1keV)$ \cite{nmasses}.
A more natural possibility is that this 
$U(1)_{Z^\prime}$ symmetry is broken at a high scale, 
which generates a large scale seesaw and naturally produces 
light neutrino masses \cite{nmasses}.
However, high scale breaking of 
the $U(1)_{Z^\prime}$ symmetry of eq. (\ref{u1zpinso10})
would naively generate effective dimension four 
proton decay mediating operators via the nonrenormalisable terms  
\beq
QLd{\cal N}\phi^n~~~~~~~~~~~~~~~~udd{\cal N}\phi^n
\label{deqfourop}
\eeq
where ${\cal N}$ and $\bar{\cal N}$ are the components of the 
Higgs fields that break $U(1)_{Z^\prime}$, and 
$\phi^n$ is a string of states that get VEVs of the order of the 
string scale. 
The induced dimension four operators are proportional to the 
$U(1)_{Z^\prime}$ symmetry breaking scale, and in the absence of additional
suppression generate proton decay at an unacceptable rate.
It is noted that in heterotic string models the problems of proton
stability and light neutrino masses are in conflict. Namely, 
the first prefers a low $U(1)_{Z^\prime}$ breaking scale, 
whereas the second works better with a high 
$U(1)_{Z^\prime}$ breaking scale. Another possibility 
is the existence of alternative gauge symmetries 
in the string vacua, which suppress the proton 
decay mediating operators but allow a high 
breaking of $U(1)_{Z^\prime}$ and therefore a 
high seesaw mass scale.
Such gauge symmetries within reach of the LHC should additionally: 
1. be anomaly free; 2. be family universal; 3. allow for quark
and lepton mass terms.
Pati proposed in ref. \cite{pati} 
that the family universal anomaly free 
combination of the flavour $U(1)$s in the model of ref. \cite{slm}
plays a role in adequately suppressing proton decay mediating operators, 
as well as allowing for suppression of left--handed neutrino masses 
via the seesaw mechanism. In ref. \cite{plboxone} it was shown
that the $U(1)$ symmetry discussed in ref. \cite{pati} must be 
broken near the string scale.
Other $U(1)$  symmetries that may play a role in suppressing
proton decay mediating operators while allowing for 
a high seesaw mass scale were discussed in refs. \cite{cfg}. 
To understand the properties of these extra $U(1)$ symmetries, 
and why they fail to materialise as low scale $Z^\prime$s, 
it is instrumental to first consider
the general structure of the free fermionic models.

\section{Free fermionic constructions}

In the free fermionic construction all the 
degrees of freedom needed to cancel the world--sheet conformal 
anomaly are represented in terms of Majorana--Weyl free fermions. 
It is important to emphasise that the fermions are free only at
a specific point in the moduli space and moving away from that 
point entails adding world--sheet Thirring interactions \cite{chang}, 
which preserve the conformal symmetry. The constructions are mathematically
equivalent to bosonic compactifications on six dimensional tori, 
with the world--sheet Thirring interactions being equivalent to 
the exact marginal deformations in the bosonic models. In four
dimensional models in the light--cone gauge the total number of 
world--sheet fermions is twenty left--moving and forty--four 
right--moving real two dimensional fermions. Eight of the 
left--moving fermions correspond to the Ramond--Neveu--Schwarz
fermions in the supersymmetric side of the ten dimensional 
heterotic--string, whereas the additional twelve correspond to 
the six left--moving compactified coordinates. Similarly, 
on the right--moving side twelve real fermions correspond to 
the six compactified dimensions. The remaining thirty--two
fermions are combined into sixteen complex fermions that
give rise to the rank sixteen gauge symmetry of the 
heterotic--string in ten dimensions. The sixty--four 
world--sheet fermions are typically denoted by
$\{\psi^{1,2}, (\chi, y, \omega)^{1,\cdots, 6}
\vert({\bar y},{\bar\omega})^{1,\cdots, 6}, 
{\bar\psi}^{1, \cdots,5},{\bar\eta}^{1,2,3},{\bar\phi}^{1,\cdots,8}
\}$, 
where ${\bar\psi}^{1,\cdots,5}$ are the Cartan generators of the
$SO(10)$ GUT group. 

Under parallel 
transport around the non--contractible loops of the torus 
representing the vacuum to vacuum amplitude, the world--sheet 
fermions pick up a phase. These transformation properties are 
summarised in sixty--four dimensional vectors. 
Invariance of the vacuum to vacuum amplitude under modular
transformations leads to a set of constraints on the
phase assignments. Summation over all the possible assignments
with appropriate phases to render a sum which is modular
invariant produces the partition function. Models in the 
free fermionic construction are therefore obtained by 
specifying a set of boundary condition basis vectors
and the one--loop summation phases in the partition function.
The resulting string vacuum may be equivalently 
specified as an orbifold of a six dimensional internal torus, 
and corresponds to compactification on some Calabi-Yau manifold
in the smooth effective field theory limit.
The moduli deformations of the 
six dimensional internal manifold are represented 
in the fermionic construction in terms of world--sheet Thirring
interactions.
The massless spectrum and interactions are obtained in the 
free fermionic formalism by applying the Generalised GSO 
projections and can be extracted in relative ease. 

The early free fermionic heterotic--string models were 
constructed by specifying a set of eight (or nine) 
basis vectors. The first five basis vectors 
consist of the so--called NAHE--set \cite{nahe}, and are common in 
all the early phenomenological models.
The gauge group at the level of the NAHE--set is 
$SO(10)\times SO(6)^3\times E_8$, with forty--eight 
multiplets in the spinorial $16$ representation of $SO(10)$. 
The $SO(10)$ symmetry is broken to one of its subgroups and the 
numbers of generations is reduced to three by adding 
three or four basis vectors to the NAHE--set. 
The phenomenological properties of the models are 
then extracted by calculating trilevel and higher 
order terms in the superpotential and by analysing 
its flat directions. 

\section{Toward string predictions}

The phenomenological three generation models may also lead to 
signatures beyond the Standard Model, that may be tested
in future experiments. It is important to note that the
actual correlation of these effects with experimental 
data will necessarily employ an effective field theory
parameterisation. Among the possibilities we 
may list: specific patterns of supersymmetry breaking,
which may be seen in forthcoming collider experiments; 
additional spacetime gauge bosons that similarly 
lead to specific collider signatures; and the 
existence of exotic matter that produces
a variety of dark matter candidates.
Recently the ATLAS and CMS experiments at the 
LHC reported possible excesses that are compatible 
with the existence of additional vector bosons
of order 2TeV \cite{atlascms}. A comprehensive analysis
of possible extra vector bosons in string 
models was undertaken in ref. \cite{guzzi}. 

However, the construction of string models that 
allow the existence of an extra $Z^\prime$, of the type 
that may be observed by the LHC experiments at the 
TeV scale, is highly non--trivial. To see
why this is the case we have to examine the patterns
of $SO(10)$ symmetry breaking induced by the 
basis vectors beyond the NAHE--set. 
The $SO(10)$ symmetry is broken to the following subgroups
by the assignment of boundary conditions $b({\bar\psi}^{1\cdots5}_{1\over2})$:

\begin{eqnarray}
1.~b\{{{\bar\psi}_{1\over2}^{1\cdots5} {~{\bar\eta}^1}~ {{\bar\eta}^2} ~{ {\bar\eta}^3}}\}&=&
\{ {1\over2}{1\over2}{1\over2}{1\over2}{1\over2}{{1\over2}} {{1\over2}} { {1\over2}}\} 
\Rightarrow ~
SU(5)\times U(1) ~\times U(1)\times U(1)\times U(1) \label{fsu5bp}\\
2.~b\{{{\bar\psi}_{1\over2}^{1\cdots5}
{~{\bar\eta}^1}~ {{\bar\eta}^2} ~{ {\bar\eta}^3}
}\}&=&\{~1 ~1 ~1 ~0 ~0~ {0} {0} {0} ~\}
\Rightarrow SO(6)\times SO(4)\times U(1)\times U(1)\times U(1). \label{so64bp}
\end{eqnarray}
The breaking 
$SO(10) ~~\rightarrow SU(3)_C\times SU(2)_L\times U(1)_C\times U(1)_L$
is obtained by combining ({\ref{fsu5bp}) and (\ref{so64bp}) in two 
separate basis vectors. The Left--Right Symmetric (LRS)
breaking pattern
%
$SO(10)\rightarrow SU(3)_C\times SU(2)_L\times SU(2)_R \times U(1)_{B-L}$
may be obtained by including two $SO(10)$ breaking basis vectors with
the first inducing the Pati--Salam breaking pattern in 
eq. (\ref{so64bp}) and the second
\begin{eqnarray}
3.~b\{{{\bar\psi}_{1\over2}^{1\cdots5}~
{{\bar\eta}^1}~ {{\bar\eta}^2} ~{{\bar\eta}^3}
}\}&=&
\{{1\over2}{1\over2}{1\over2}~0~0~
{{1\over2}}{{1\over2}} {{1\over2}}\}\Rightarrow 
\label{su3122b}\\
& &~~~~~~~~~~~~~~~~~~
SU(3)_C \times U(1)_C \times SU(2)_L\times SU(2)_R \times 
{U(1)}\times  {U(1)}\times {U(1)} \nn
\end{eqnarray}
A key difference between the 
$SU(5)\times U(1)$, 
$SO(6)\times SO(4)$
and $SU(3)\times SU(2)\times U(1)^2$ cases, which are obtained 
by using the boundary condition assignments in eqs. (\ref{fsu5bp})
and (\ref{so64bp}), versus the LRS models 
is with respect to the charges of the Standard Model family states
%
under the three $U(1)$ symmetries, $U(1)_{1,2,3}$, corresponding 
to the world--sheet complex fermions ${\bar\eta}^{1,2,3}$.
In the first class of models 
$Q_{U(1)_j} (16 = \{Q, L, U, D, E, N\}) ~=~ + {1\over2}$, 
{\it i.e.} all the states
in the $16$ $SO(10)$ spinorial representation in a given 
twisted plane have common charge $\pm1/2$. 
Therefore, in these models $U(1)_{1,2,3}$,
as well as the the family universal combination 
$U(1)_\zeta=U(1)_1+U(1)_2+U(1)_3$, are anomalous.
In the LRS models on the other hand, the charges 
differ between the left-- and right--handed states, with 
\beqn
& & Q_{U(1)_j} (~~~~~~~~~~~~~~~Q_L~~~,~~~ L_L~~~~~~~~~~~~~~~) 
~~~=~ - {1\over2}~,\label{lrscal}\\
& & Q_{U(1)_j} (~Q_R = \{U,D\}~~, ~~L_R=\{E,N\}~) ~=~ + {1\over2}~.
\label{lrscar}
\eeqn
Consequently, in these models the $U(1)_{1,2,3}$ symmetries as well 
as the family universal combination $U(1)_\zeta$ are anomaly free.
Three generation left--right symmetric string derived models
were presented in ref. \cite{lrs}. Some of the models presented 
contain untwisted Higgs bi--doublets that can 
be used to generate quasi--realistic fermion mass spectrum
from renormalisable and non--renormalisable superpotential terms.

For a $U(1)$ symmetry to remain unbroken down to low 
scales it must be anomaly free. To be viable at scales 
within reach of the LHC it must also be family universal. 
The family universal combination $U(1)_\zeta$ satisfies 
the two requirements. To study its signatures at 
low scales we build a string inspired model \cite{viraf} with the 
charge assignments in eqs. (\ref{lrscal}, \ref{lrscar}). 
Considering only the Standard Model matter states leads
to a mixed $SU(2)_L^2\times U(1)_\zeta$ anomaly. The string derived
models are anomaly free and contain extra states that cancel the
anomaly. To construct a model free of all anomalies 
$SU(2)_{L/R}$ doublets are added to the model. To facilitate
gauge coupling unification we may also add colour triplets. 
We impose that the charge assignments in the string
inspired model are compatible with the charges in the 
string derived models. Extrapolating the gauge coupling from 
the GUT unification scale to the electroweak scale and 
imposing the experimental values $\alpha_S(M_Z)\approx0.1$ 
and $\sin^2\theta(M_Z)\approx0.231$ yields the following 
hierarchy of scales \cite{viraf} 
 $$M_{SUSY}\approx 1{\rm TeV};~~
                 M_{Z^\prime}> 10^8{\rm GeV};~~ 
                 M_D> 10^{12}{\rm GeV};~~
                 M_R\approx M_{string}, $$
where $M_R$ is the symmetry breaking scale of $SU(2)_R$, 
which induces the seesaw mechanism and the suppression of 
left--handed neutrino masses. It is noted that compatibility of 
string gauge coupling unification with the gauge coupling parameters 
at the electroweak scale requires that $M_{Z^\prime}$ is heavier
than $10^8$GeV. Hence, a $Z^\prime$ at the TeV scale in this 
string inspired model is incompatible with the gauge coupling data. 
The root of the discrepancy can be seen to arise from the 
fact that the $Z^\prime$ charges do not admit the $E_6$ embedding
in this model. The contrast between this model versus models 
that admit the $E_6$ embedding can be seen by examining the 
contributions of the intermediate gauge and matter 
thresholds to $\sin^2\theta_W(M_Z)$ and $\alpha_3(M_Z)$.
In the case of the charge assignments in the LRS string 
inspired model the threshold corrections from intermediate gauge 
and matter scales are given by 
\begin{eqnarray}
\delta\left(\sin^2\theta_W(M_Z)\right)_{\mbox{\tiny{I.T.}}} & = &
\frac{1}{2\pi}\frac{k_1\alpha}{1+k_1}
\left(
\frac{12}{5}\log\frac{M_S}{M_R}-
\frac{24}{5}\log\frac{M_S}{M_{Z^\prime}}-
\frac{2n_D}{5}\log\frac{M_S}{M_D}\right), \nn\\
\delta\left(\alpha_3(M_Z)\right)_{\mbox{\tiny{I.T.}}} & = &
\frac{1}{2\pi}
\left(
\frac{3}{2}\log\frac{M_S}{M_R}-
9\log\frac{M_S}{M_{Z^\prime}}+
\frac{3n_D}{4}\log\frac{M_S}{M_D}\right), \label{deltalrs}
\end{eqnarray}
whereas in the case of models that admit the $E_6$ embedding 
the threshold corrections are given by 
\begin{eqnarray}
\delta\left(\sin^2\theta_W(M_Z)\right)_{\mbox{\tiny{I.T.}}}  &= &
\frac{1}{2\pi}\frac{k_1\alpha}{1+k_1}
\left(
\frac{12}{5}\log\frac{M_S}{M_R}+
\frac{6}{5}\log\frac{M_S}{M_H}-
\frac{6}{5}\log\frac{M_S}{M_D}\right),\nn\\
\delta\left(\alpha_3(M_Z)\right)_{\mbox{\tiny{I.T.}}}& = &
\frac{1}{2\pi}
\left(
\frac{3}{2}\log\frac{M_S}{M_R}-
\frac{9}{4}\log\frac{M_S}{M_H}+
\frac{9}{4}\log\frac{M_S}{M_D}\right).
\label{deltae6e}
\end{eqnarray}
Eqs. (\ref{deltalrs}) and (\ref{deltae6e}) demonstrate the cancellation 
between thresholds corrections of 
the doublets and triplets in the models that admit 
the $E_6$ embedding, which is not the case in the string inspired LRS models, 
as seen from eq. (\ref{deltalrs}).
Compatibility of heterotic--string gauge coupling 
unification with the gauge coupling data at the electroweak 
scale therefore favours string models that admit the $E_6$ 
embedding of the $U(1)_\zeta$ charges. As discussed above
in many of the string derived models, including 
the flipped $SU(5)$ \cite{fsu5}, the Pati--Salam \cite{alr}
and the standard--like models \cite{slm}, $U(1)_\zeta$ is anomalous, 
which results from the symmetry breaking pattern
$E_6\rightarrow SO(10)\times U(1)_\zeta$, induced by a
GGSO projection at the string scale. Construction 
of models with $U(1)_\zeta$ charges that admit an $E_6$ embedding
and in which $U(1)_\zeta$ is anomaly free can be pursued 
in two directions. The first is to construct string models 
in which the symmetry breaking pattern 
$E_6\rightarrow SO(10)\times U(1)_\zeta$ is not present. 
To understand how this is achieved it is instrumental 
to examine how the gauge symmetry is generated 
in the free fermionic models. Observable space--time vector
bosons in the relevant models are obtained from two sectors. 
The first is the untwisted Neveu--Schwarz--sector (NS--sector) that 
produces vector bosons in the adjoint representation of $SO(2n)$ 
groups, with $n\leq8$. The second is the $x$--sector \cite{fmmckm}
that produces states in the spinorial representation of $SO(2n)$. 
In the case of the $E_6$ symmetry, the NS--sector produces the 
vector bosons in the adjoint representation of $SO(10)\times U(1)$,
whereas the $x$--sector produces vector bosons in the 
$16_{+1}\oplus{\overline{16}}_{-1}$ representation of the 
$SO(10)\times U(1)$ gauge group.
The breaking in the string models of $E_6$ to $SO(10)\times U(1)_\zeta$
is obtained by projecting the $16_{+1}\oplus{\overline{16}}_{-1}$ 
spinorial states from the $x$--sector. Maintaining the 
$E_6$ embedding of the $U(1)_\zeta$ charges, while keeping 
as an anomaly free symmetry can therefore be obtained 
by keeping the states from the $x$--sector in the 
massless spectrum, and inducing alternative $E_6$ symmetry
breaking patterns. An example of models in this class 
includes the $SU(6)\times SU(2)$ heterotic--string
derived models of ref. \cite{su62}. The caveat is that, 
given the available Standard Model singlets in the 
spectrum of the heterotic--string model, it is not possible
to break the $SU(6)\times SU(2)$ to $SU(3)_C\times SU(2)_L\times U(1)_Y\times
U(1)_{Z^\prime}$, {\it i.e.} the extra $Z^\prime$ is necessarily broken 
at a high scale by the 
VEVs of the Standard Model singlets. The available Standard Model
singlets are the $SO(10)$ singlet from the 27 representation
of $E_6$ and the second is the Standard Model singlet 
in the spinorial 16 representation of $SO(10)$. A VEV for either
of these fields leaves an unbroken non--Abelian extension of the 
Standard Model, and reduction to the non--Abelian content 
of the Standard Model uses both VEVs. Construction of string models
with $SU(4)\times SU(2)_L\times SU(2)_R\times U(1)_{\zeta^\prime}$ 
and  $SU(3)\times SU(2)_L\times U(1)_Y\times U(1)_{Z^\prime}$ 
was discussed in \cite{viraf} and \cite{afp}, respectively. 
In both cases the NS--sector generates a subgroup of the 
gauge symmetry, which is enhanced by states
from the $x$--sector. Both cases would allow for an
extra $Z^\prime$ at low scales. However, concrete three generation
string models that realise this construction were 
not presented in \cite{viraf,afp}. 

\section{Spinor--vector duality} 

An alternative method to build string models with anomaly free 
$U(1)_\zeta$ is to utilise the spinor--vector duality discovered 
in the classification of $SO(10)$ heterotic--string models
\cite{classification}.
The spinor--vector duality is a symmetry in the space of heterotic--string
vacua under the exchange of the total number of $16\oplus {\overline{16}}$
spinorial and anti--spinorial representations of $SO(10)$, with the 
total number of vectorial $10$ representations \cite{svduality}. 
The statement is that for a given vacuum with a \#1
of $16\oplus {\overline{16}}$ representations, and a \#2
of vectorial $10$ representations, there exist another vacuum 
in which the two are interchanged. The duality operates in 
the space of $Z_2\times Z_2$ heterotic--string in which the 
$(2,2)$ world--sheet supersymmetry is broken to $(2,0)$. 
In fact, it is possible to understand the origin of the duality 
from its origin in the $(2,2)$ vacua. In those cases 
the $SO(10)\times U(1)$ symmetry is enhanced to $E_6$. 
The $27$ representation of $E_6$ decomposes under 
$SO(10)\times U(1)$ as $27=16_{1/2}+ 10_{-1}+ 1_{+2}$,
whereas the ${\overline{27}}$ decomposes as 
$27={\overline{16}}_{-1/2}+ 10_{+1}+ 1_{-2}$. 
Thus, in the case of vacua with $(2,2)$ world--sheet 
supersymmetry the total number of 
$16\oplus {\overline{16}}$ representations is equal
to the total number of $10$ representations, {\it i.e.}
\#1$=$\#2. This case therefore corresponds to the enhanced symmetry,
self--dual point under the spinor--vector exchange.
This is similar to the case of $T$--duality \cite{tduality} 
where the symmetry at the self--dual point is enhanced from
$U(1)^2$ to $SU(2)^2$. At the level of the $(2,2)$ 
there exist a spectral flow operator, on the bosonic 
side of the heterotic--string, that exchanges between
the components of the $SO(10)\times U(1)$ representations
inside the $27$ representation of $E_6$. This spectral 
flow operator is a generator of the world--sheet $N=2$ 
supersymmetry on the bosonic side of the 
heterotic--string. It is similar to the 
spectral flow operator on the fermionic side, which 
acts as the space--time supersymmetry generator.
When the world--sheet supersymmetry is broken 
on the bosonic side of the heterotic--string by 
a Wilson line the $E_6$ symmetry is broken to $SO(10)\times U(1)$. 
One has a choice of using a Wilson line that leaves 
a \#1 of massless $16\oplus {\overline{16}}$ representations,
and a \#2 of vectorial $10$ representations, 
or a choice of a second Wilson line, which exchanges
the two numbers \cite{florakis}. The map between the 
two Wilson lines, or between the two resulting string 
vacua, is induced by the spectral flow operator
on the bosonic side of the heterotic--string.
Hence, the spinor--vector duality results from the breaking 
of the $(2,2)$ world--sheet supersymmetry to $(2,0)$ 
and is induced by the spectral flow operator
on the bosonic side of the heterotic--string
\cite{florakis}.
The spinor--vector duality is a remarkable property in the space
of $Z_2$ and $Z_2\times Z_2$ heterotic--string vacua, 
akin to mirror symmetry \cite{mirror}.
It indicates a global structure underlying the entire space 
of solutions, and may be a manifestation of a deeper 
mathematical structure underlying these compactifications. 
The picture may be extended  to string theories with 
interacting world--sheet CFTs, {\it e.g.} to Gepner models \cite{gepner}, 
albeit with slightly more intricate relations \cite{panos}.


As noted above the self--dual vacua under the spinor--vector duality map are
those in which the total number of $16\oplus {\overline{16}}$ is
equal to the total number of $10$ representations. This self--duality
property is realised when the gauge symmetry is enhanced to
$E_6$. In this case $U(1)_\zeta$ is anomaly free by virtue of
its embedding in $E_6$. However, there is a class of self--dual 
vacua with equal number of $16\oplus {\overline{16}}$ and $10$ 
representations in which the symmetry is not enhanced to $E_6$.
This is possible if the $16$ and $10$ components are 
obtained at different fixed points of the $Z_2\times Z_2$
orbifold. That is, obtaining a spinorial $16$ multiplet
and a vectorial $10$ multiplet at the same fixed point
would necessarily imply enhancement of the gauge
symmetry to $E_6$. However, if they are obtained at 
different fixed points we may have models in 
which their number is equal, {\it i.e.} which 
are self--dual with respect to the spinor--vector
duality, but in which the gauge symmetry is not enhanced 
to $E_6$. In such vacua $U(1)_\zeta$ may be anomaly free 
because the chiral spectrum comes in complete $E_6$ 
multiplets. However, the $E_6$ is not manifested 
in the low energy effective field theory. 

The spinor--vector duality was initially found ``empirically''
by the classification of free fermionic models with $SO(10)$ 
GUT symmetry, and proven in terms of the GGSO 
projection coefficients. It was further demonstrated
in terms of discrete torsions of the $Z_2^2$ and $Z_2^3$ orbifolds, 
and as a map between two Wilson lines, induced by the 
spectral flow operator, as discussed above. 

In a model that may give rise to a low scale $Z^\prime$
the $SO(10)$ symmetry is necessarily broken. However, 
we may seek models that preserve the self--duality 
property at the $SO(10)$ level. In such models the 
chiral spectrum will reside in complete $E_6$ 
multiplets, decomposed under the effective unbroken gauge
symmetry. In such models $U(1)_\zeta$ may be anomaly free 
and be a component of a low scale $Z^\prime$. 
To ``troll'' a model with these features we use
the classification methodology developed in ref. 
\cite{classification, psclassi, fsuclassi},
which is outlined in the next section.  


\section{The classification methodology}

The early three generation models in the free fermionic formulation
\cite{fsu5, slm, alr} consisted of isolated examples, 
and were obtained, in a sense, by a straw of good fortune.
A more systematic approach was developed over the 
past two decades, which entails the scanning 
of large spaces of fermionic $Z_2\times Z_2$ 
orbifold compactifications, of the order 
of $10^{15}$ vacua. The method was developed 
in \cite{typeIIbclass} for the classification 
of type IIB superstrings and extended in 
\cite{classification, psclassi, fsuclassi}
for the classification of heterotic string vacua with various 
subgroups of an $SO(10)$ GUT group. In this method 
the set of boundary condition basis vectors is fixed and
the enumeration of the models is obtained by varying 
the Generalised GSO (GGSO) phases.
The set of basis vectors used is given
by a set of thirteen basis vectors
$
B=\{v_1,v_2,\dots,v_{13}\}. 
$
The first twelve basis vectors are shown in eq. 
(\ref{so10basis}),
\begin{eqnarray}
v_1=1&=&\{\psi^\mu,\
\chi^{1,\dots,6},y^{1,\dots,6}, \omega^{1,\dots,6}| \nonumber\\
& & ~~~\bar{y}^{1,\dots,6},\bar{\omega}^{1,\dots,6},
\bar{\eta}^{1,2,3},
\bar{\psi}^{1,\dots,5},\bar{\phi}^{1,\dots,8}\},\nonumber\\
v_2=S&=&\{\psi^\mu,\chi^{1,\dots,6}\},\nonumber\\
v_{2+i}=e_i&=&\{y^{i},\omega^{i}|\bar{y}^i,\bar{\omega}^i\}, \
i=1,\dots,6,\nonumber\\
v_{9}=z_1&=&\{\bar{\phi}^{1,\dots,4}\},\label{so10basis}\\
v_{10}=z_2&=&\{\bar{\phi}^{5,\dots,8}\},\nonumber\\
v_{11}=b_1&=&\{\chi^{34},\chi^{56},y^{34},y^{56}|\bar{y}^{34},
\bar{y}^{56},\bar{\eta}^1,\bar{\psi}^{1,\dots,5}\},\nonumber\\
v_{12}=b_2&=&\{\chi^{12},\chi^{56},y^{12},y^{56}|\bar{y}^{12},
\bar{y}^{56},\bar{\eta}^2,\bar{\psi}^{1,\dots,5}\},\nonumber
\end{eqnarray}
where the fermions appearing in the curly brackets in eq. (\ref{so10basis})
are periodic, whereas those that do not appear are antiperiodic.
The set of twelve basis vectors 
appearing in (\ref{so10basis}) is identical to the 
set used to generate the $SO(10)$ models. The first ten 
basis vectors $\{1, S, z_1, z_2, e_i\} $,  with $i=1, \dots, 6$, 
generate vacua that preserve $N=4$ space--time supersymmetry. 
The subsequent two basis vectors, $b_1$ and $b_2$, 
correspond to the $Z_2\times Z_2$ orbifold twistings and 
break $N=4$ space--time supersymmetry to $N=2$ and $N=1$.

The thirteenth vector in our basis breaks the $SO(10)$ symmetry to
a subgroup. The vector given by 
\beq
v_{13}=\alpha = \{\bar{\psi}^{4,5},\bar{\phi}^{1,2}\}, \label{psvector}
\eeq
is used to generate the Pati--Salam subgroup \cite{psclassi}, 
whereas the basis vector
\beq
v_{13}=\beta =\{\overline{\psi}^{1,\dots,5}=\textstyle\frac{1}{2},
\overline{\eta}^{1,2,3}=\textstyle\frac{1}{2},
\overline{\phi}^{1,2} = \textstyle\frac{1}{2}, 
\overline{\phi}^{3,4} = \textstyle\frac{1}{2},
\overline{\phi}^{5}=1,\overline{\phi}^{6,7}=0,
\overline{\phi}^{8}=0\,\},\label{fsu5vector}
\eeq
is used in the case of the flipped $SU(5)$ models \cite{fsuclassi}. 
We note that the 
models contain many sectors that may enhance the four dimensional
gauge group. In the model that we seek to obtain here we impose
that all the additional space--time vector bosons that arise in
these additional sectors are projected out from the massless
spectrum. The self--dual model that we seek to present here
is obtained by using the basis vector in eq. (\ref{psvector}). 
Hence, the $SO(10)$ symmetry in this model is broken to the 
Pati--Salam subgroup. The one--loop GGSO phases 
in the partition function are given by a $13\times 13$ matrix
$$%
c{{v_i}\atopwithdelims[] {v_j}} = %
\bordermatrix{%
         &1   &  S  &  e_1  &   e_2   &  e_3   &  e_4  &   e_5  &   e_6%
&  z_1   &  z_2   &  b_1  &   b_2 & \alpha\cr%
   1   & -1   &  -1 &  \pm  &  \pm  &  \pm  &  \pm  &  \pm   &  \pm  & \pm  &%
 \pm   & \pm  & \pm & \pm\cr%
   S~   &   &    & -1 & -1 & -1 & -1 & -1 & -1 & -1 & -1 &  1 & 1 & -1\cr%
  e_1~  &   &    &    &\pm &\pm &\pm &\pm &\pm &\pm &\pm &\pm &\pm&\pm\cr%
  e_2~  &   &    &    &    &\pm &\pm &\pm &\pm &\pm &\pm &\pm &\pm&\pm\cr%
  e_3~  &   &    &    &    &    &\pm &\pm &\pm &\pm &\pm &\pm &\pm&\pm\cr%
  e_4~  &   &    &    &    &    &    &\pm &\pm &\pm &\pm &\pm &\pm&\pm\cr%
  e_5~  &   &    &    &    &    &    &    &\pm &\pm &\pm &\pm &\pm&\pm\cr%
  e_6~  &   &    &    &    &    &    &    &    &\pm &\pm &\pm &\pm&\pm\cr%
  z_1~  &   &    &    &    &    &    &    &    &    &\pm &\pm &\pm&\pm\cr%
  z_2~  &   &    &    &    &    &    &    &    &    &    &\pm &\pm&\pm\cr%
  b_1~  &   &    &    &    &    &    &    &    &    &    &    &\pm&\pm\cr%
  b_2~  &   &    &    &    &    &    &    &    &    &    &    & -1&\pm\cr%
\alpha~ &   &    &    &    &    &    &    &    &    &    &    &   &   \cr%
  },%
$$%
where only the entries above the diagonal are independent, whereas 
those on and below the diagonal are determined by modular invariance.
The entries in the second row are fixed by requiring that the 
models preserve $N=1$ space--time supersymmetry, 
and the phase $c{{b_1}\atopwithdelims[] {b_2}}$
only affects the overall chirality, and is therefore fixed. This 
leaves a priori a space of $2^{66}$ vacua.
Below we introduce the notation
$$c{{vi}\atopwithdelims[] {v_j}}~=~{\rm exp}[i\pi(v_i|v_j)],$$
which is instrumental for the analysis. 

The classification methodology facilitates systematic extraction 
of the entire twisted massless spectrum. The untwisted spectrum
is common in all the models by demanding that all enhanced 
symmetries are projected out. Focusing on the Pati--Salam
class of models, the observable gauge symmetry is 
then $SO(6)\times SO(4)\times U(1)^3$. 
The chiral generations, for example,
then arise from the twisted sectors 
\begin{eqnarray} \label{chiralgen}
B_{pqrs}^{(1)}&=& S + b_1 + p e_3+ q e_4 + r e_5 + s e_6 \nonumber\\
&=&\{\psi^\mu,\chi^{12},(1-p)y^{3}\bar{y}^3,p\omega^{3}\bar{\omega}^3,
(1-q)y^{4}\bar{y}^4,q\omega^{4}\bar{\omega}^4, \nonumber\\
& & ~~~(1-r)y^{5}\bar{y}^5,r\omega^{5}\bar{\omega}^5,
(1-s)y^{6}\bar{y}^6,s\omega^{6}\bar{\omega}^6,\bar{\eta}^1,\bar{\psi}^{1..5}\}
\\
B_{pqrs}^{(2)}&=& S + b_2 + p e_1+ q e_2 + r e_5 + s e_6
\label{chiralspinorials2}
\nonumber\\
B_{pqrs}^{(3)}&=& S + b_3 + p e_1+ q e_2 + r e_3 + s e_4 \nonumber
\end{eqnarray}
where $p,q,r,s=0,1$; $b_3=b_1+b_2+x=1+S+b_1+b_2+\sum_{i=1}^6 e_i+\sum_{n=1}^2
z_n$ and
$x$ is given by $x=\{{\bar\psi}^{1,\cdots,5},{\bar\eta}^{1,2,3}\}$.
These sectors produce $16$ and $\overline{16}$
representations of $SO(10)$ decomposed under
$SO(6)\times SO(4)\equiv SU(4)\times SU(2)_L\times SU(2)_R$,
\beqn
\textbf{16} = & \textbf{(4,~2,~1)} + \textbf{(}\bar{\textbf{4}}\textbf{, 1, 2)}
\nonumber\\
\overline{\textbf{16}} = &\textbf{(}\bar{\textbf{4}}\textbf{, 2, 1)} +
\textbf{(4,~1,~2)} \nonumber
\eeqn
Each of the 48 sectors can give rise at most to one state.
The GGSO projections can be recast as algebraic equations. 
For example, for the chiral matter arising in the sectors
above, we can write projector equations in terms 
of the GGSO phases given in matrix form  
$\Delta^{i}W^{i} = Y^{i}$.
\begin{eqnarray}
\left(
\begin{array}{cccc}
\oo{e_1}{e_3}&\oo{e_1}{e_4}&\oo{e_1}{e_5}&\oo{e_1}{e_6}\\
\oo{e_2}{e_3}&\oo{e_2}{e_4}&\oo{e_2}{e_5}&\oo{e_2}{e_6}\\
\oo{z_1}{e_3}&\oo{z_1}{e_4}&\oo{z_1}{e_5}&\oo{z_1}{e_6}\\
\oo{z_2}{e_3}&\oo{z_2}{e_4}&\oo{z_2}{e_5}&\oo{z_2}{e_6}
\end{array}
\right)
\left(
\begin{array}{c}
p\\
q\\
r\\
s\\
\end{array}
\right)
=
\left(
\begin{array}{c}
\oo{e_1}{b_1}\\
\oo{e_2}{b_1}\\
\oo{z_1}{b_1}\\
\oo{z_2}{b_1}
\end{array}
\right) \nonumber
\ea
\vspace{-0.2cm}
\ba
\left(
\begin{array}{cccc}
\oo{e_3}{e_1}&\oo{e_3}{e_2}&\oo{e_3}{e_5}&\oo{e_3}{e_6}\\
\oo{e_4}{e_1}&\oo{e_4}{e_2}&\oo{e_4}{e_5}&\oo{e_4}{e_6}\\
\oo{z_1}{e_1}&\oo{z_1}{e_2}&\oo{z_1}{e_5}&\oo{z_1}{e_6}\\
\oo{z_2}{e_1}&\oo{z_2}{e_2}&\oo{z_2}{e_5}&\oo{z_2}{e_6}
\end{array}
\right)
\left(
\begin{array}{c}
p\\
q\\
r\\
s\\
\end{array}
\right)
=
\left(
\begin{array}{c}
\oo{e_3}{b_2}\\
\oo{e_4}{b_2}\\
\oo{z_1}{b_2}\\
\oo{z_2}{b_2}
\end{array}
\right)
\ea
\vspace{-0.2cm}
\ba
\left(
\begin{array}{cccc}
\oo{e_5}{e_1}&\oo{e_5}{e_2}&\oo{e_5}{e_3}&\oo{e_5}{e_4}\\
\oo{e_6}{e_1}&\oo{e_6}{e_2}&\oo{e_6}{e_3}&\oo{e_6}{e_4}\\
\oo{z_1}{e_1}&\oo{z_1}{e_2}&\oo{z_1}{e_3}&\oo{z_1}{e_4}\\
\oo{z_2}{e_1}&\oo{z_2}{e_2}&\oo{z_2}{e_3}&\oo{z_2}{e_4}
\end{array}
\right)
\left(
\begin{array}{c}
p\\
q\\
r\\
s\\
\end{array}
\right)
=
\left(
\begin{array}{c}
\oo{e_5}{b_3}\\
\oo{e_6}{b_3}\\
\oo{z_1}{b_3}\\
\oo{z_2}{b_3}
\end{array}
\right) \nonumber
\end{eqnarray}
The total number of states then correspond to the total number 
of solutions for $pqrs$, which is determined by the rank of 
the matrices $\Delta^i$ 
relative to the rank of the augmented matrices ($\Delta^i, Y^i$)
\cite{classification}. Similarly, the chirality of the states, 
or the charges of the periodic fermions with respect to the 
$U(1)$ generators of the Cartan subalgebra, can be written
as generic expressions in terms of the GGSO phases. The 
vectorial $SO(10)$ representations are obtained from the sectors
$B^{(i)}_{pqrs}+x$ and similar algebraic expressions are written. 
The entire massless spectrum can be extracted in a similar 
fashion, and coded in a computer program. This enables 
exploration of the entire space of string vacua and 
analysis of their massless spectra. In some cases
the classification of the complete space of solutions
is not feasible 
and a statistical algorithm is developed
by generating random choices of GGSO projection phases.
In figures \ref{figure1} and \ref{figure2} we display 
some of the results in the classification of the 
Pati--Salam and flipped $SU(5)$ models, respectively. 
\newpage

\begin{figure}[!]
	\centering
	\includegraphics[width=100mm]{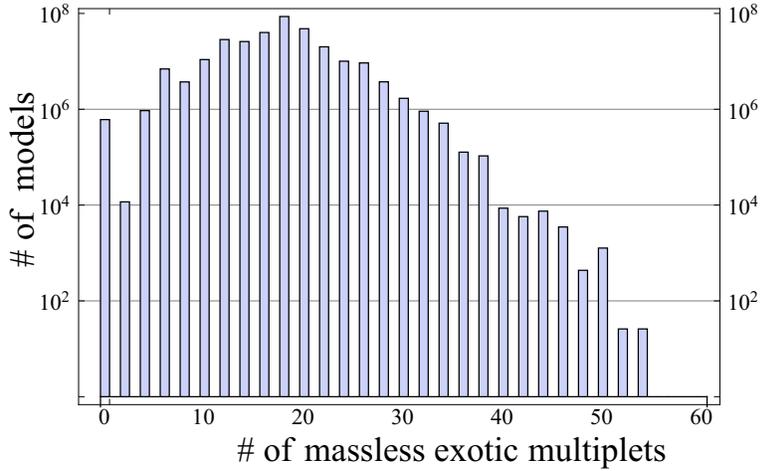}
	\caption{\emph{
Number of 3--generation models versus total number of exotic multiplets
 in a random sample of $10^{11}$ Pati-Salam configurations.
}
}
	\label{figure1}
\end{figure}

The striking observations is that while the Pati--Salam 
case admits exophobic three generation models in the flipped
$SU(5)$ case exophobic models with an odd number of generations
do not appear. While positive confirmation, like the existence
of a three generation exophobic model in the Pati--Salam case, 
is easy to verify, understanding their absence in the flipped
$SU(5)$ case is more challenging. For example, it may merely 
reflect their absence in the particular space sampled.
Nevertheless, the results shown in figures \ref{figure1} 
and \ref{figure2} exemplify the power of the classification
methodology in extracting properties of large spaces of 
string vacua. Using the random generation of GGSO phases 
and imposing some prior criteria we can ``fish'' 
out models with desirable characteristics, provided 
that the statistical sample of the models with the 
required properties is not too small.

\section{``Trolling'' a model}

A set of phases giving rise to a model with the desired properties 
is displayed in eq. (\ref{BigMatrix}). As advertised 
the observable gauge symmetry in this model is
$SO(6)\times SO(4)\times U(1)_{1,2,3}$ and the family universal 
combination,
$U(1)_\zeta=U(1)_1+U(1)_2+U(1)_3$,
is anomaly free. 
The complete massless spectrum, as well as the cubic level 
superpotential are derived in ref. \cite{zpsmodel}. 
The massless chiral spectrum in the model appears in complete 
$27$ representations of $E_6$ decomposed 
under the effective observable gauge group, {\it i.e.} 
the chiral spectrum is self--dual under the spinor--vector duality.
The model contains three chiral generations, as well as the required
heavy and light Higgs states to produce a realistic fermion
mass spectrum. Furthermore, the model admits a top 
quark mass term at the cubic level of the superpotential, 
{\it i.e. } $\lambda_t\sim 1$ in the model. 
A VEV of the heavy Higgs field that breaks the Pati--Salam 
symmetry to the Standard Model along flat directions leaves the 
unbroken combination 
\beq
U(1)_{Z^\prime}~~=~~{1\over5}~U(1)_C~~-~~{1\over5}~U(1)_L-U_\zeta.   
\label{u1zpsdm}
\eeq
This $U(1)$ symmetry may remain unbroken down to low scales as it
is anomaly free in this model. 

\beq \label{BigMatrix}  (v_i|v_j)\ \ =\ \ \bordermatrix{
      & 1& S&e_1&e_2&e_3&e_4&e_5&e_6&b_1&b_2&z_1&z_2&\alpha\cr
 1    & 1& 1&  1&  1&  1&  1&  1&  1&  1&  1&  1&  1&      1\cr
S     & 1& 1&  1&  1&  1&  1&  1&  1&  1&  1&  1&  1&      1\cr
e_1   & 1& 1&  0&  0&  0&  0&  0&  0&  0&  0&  0&  0&      1\cr
e_2   & 1& 1&  0&  0&  0&  0&  0&  1&  0&  0&  0&  1&      0\cr
e_3   & 1& 1&  0&  0&  0&  1&  0&  0&  0&  0&  0&  1&      1\cr
e_4   & 1& 1&  0&  0&  1&  0&  0&  0&  0&  0&  1&  0&      0\cr
e_5   & 1& 1&  0&  0&  0&  0&  0&  1&  0&  0&  0&  1&      1\cr
e_6   & 1& 1&  0&  1&  0&  0&  1&  0&  0&  0&  1&  0&      0\cr
b_1   & 1& 0&  0&  0&  0&  0&  0&  0&  1&  1&  0&  0&      0\cr
b_2   & 1& 0&  0&  0&  0&  0&  0&  0&  1&  1&  0&  0&      1\cr
z_1   & 1& 1&  0&  0&  0&  1&  0&  1&  0&  0&  1&  1&      0\cr
z_2   & 1& 1&  0&  1&  1&  0&  1&  0&  0&  0&  1&  1&      0\cr
\alpha& 1& 1&  1&  0&  1&  0&  1&  0&  1&  0&  1&  0&      1\cr
  }
\eeq

We note that maintaining the $U(1)_{Z^\prime}$ symmetry in eq. 
(\ref{u1zpsdm}) unbroken down to low scales, requires 
the existence of additional matter states down to the 
$U(1)_{Z^\prime}$ breaking scale. Existence 
of this $U(1)$ symmetry at this scale will be accompanied 
by additional states with specific Standard Model 
and $U(1)_{Z^\prime}$ charges, which are mandated 
by anomaly cancellation and are compatible with the
charge assignments in the string model. 
The string models give
rise to a variety of Standard Model extensions, and each case 
is accompanied by specific additional states, which leads to 
unique signatures \cite{guzzi}. 

\begin{figure}[!]
	\centering
	\includegraphics[width=100mm]{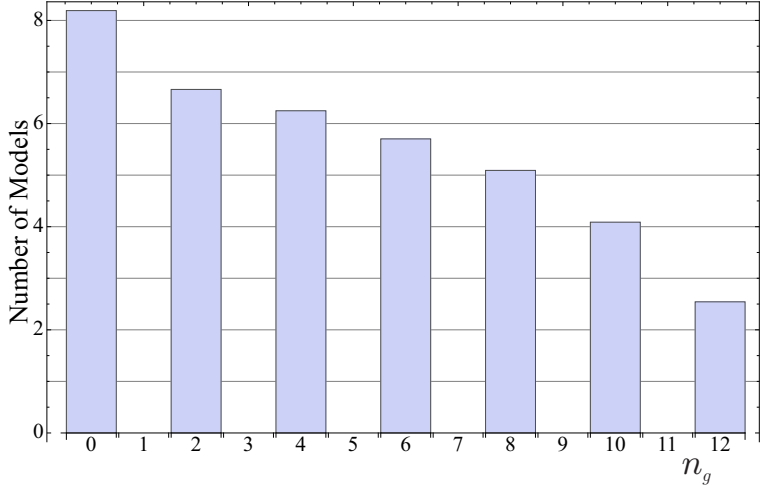}
	\caption{\emph{
Number of exophobic models versus the number of generations 
in a random sample of $10^{12}$ flipped $SU(5)$ configurations.
}
}
	\label{figure2}
\end{figure}

An additional type of states that are of interest in this model 
are states that are $SO(10)$ singlets with non--standard 
$U(1)_\zeta$ charges. Such states are standard states 
with respect to the Standard Model charges, but are exotic 
states with respect to $E_6$. They arise from 
the breaking of $E_6$ in the string model by Wilson
lines, and fall into the general category of Wilsonian
matter states discussed in ref. \cite{ssr}. 
They may play the role of dark matter candidates. 
Breaking the $U(1)_{Z^\prime}$ symmetry with states 
carry the standard conventional $E_6$ charges, 
leaves a remnant discrete symmetry that forbids
the decay of the lightest exotic state to the 
lighter states \cite{ssr}. The exotic states 
in this model are particularly appealing 
because they are Standard Model, and in
fact $SO(10)$, singlets.

\section{Conclusions}

The available observational data at accessible energy scales 
favour the high scale unification of the gauge interactions
and matter representations in Grand Unified Theories.
Gravity is left out of this picture, 
and it is anticipated that may of the properties 
of the Standard Model particles, like the origin of their
masses, can only be determined if gravity is incorporated
into the fold. However, point quantum field theories 
have not led to a consistent synthesis with gravity, 
and it may well be that consistent unification of gravity
with the gauge interactions requires a departure from the 
point particle idealisation. Contemporary string theories
therefore provide a concrete framework to explore the 
unification of gravity and the gauge interactions. 

In this paper we discussed the construction of string derived 
models that allow for the existence of an extra $U(1)$ symmetry
at low scales. While the phenomenological analysis 
of such symmetries received ample attention in the literature, 
the construction of heterotic--string models that allow for a 
light $Z^\prime$ gauge boson has proven to be problematic. 
One of the main difficulties being that the desired symmetries 
tend to be anomalous in the string derived models and
are broken at the high scale. 
For this purpose we utilised the classification methodology 
developed in the free fermionic formulation. 
We used the self--duality property under 
the spinor--vector duality to construct a string model 
in which the $E_6$ symmetries are anomaly free. This is 
obtained because due to the self--duality property, 
the chiral spectrum arises in complete 
$27$ representations of $E_6$. However, 
the remarkable point is that the gauge symmetry 
is not enhanced to $E_6$, but remains a subgroup of
anomaly free $SO(10)\times U(1)_\zeta$. This is possible because
the different components in the $27$ representations are obtained 
from different fixed points of the $Z_2\times Z_2$ orbifold. 
It is further noted that while our exemplary model is free 
of exotic fractionally charged massless states, 
it contains states that carry exotic states 
with respect to $U(1)_\zeta$, which are natural 
dark matter candidates. It will be of further
interest to explore whether similar characteristics 
can be obtained in other classes of 
heterotic--string compactifications \cite{others}. 

\section{Acknowledgments} 

AEF would like to thank the University of Oxford and CERN
for hospitality.
AEF is supported in part by STFC under contract ST/L000431/1.
This research has been co-financed by the European Union 
(European Social Fund - ESF)
and Greek
national funds through the Operational 
Program ``Education and Lifelong Learning" 
of the National Strategic Reference 
Framework (NSRF) - Research Funding Program: 
THALIS Investing in the society of knowledge 
through the European Social Fund.


\newpage

\begin{thebibliography}{99}
\bibitem{fff} H.\ Kawai, D.C.\ Lewellen, and S.H.-H.\ Tye,
					\NPB{288}{1987}{1};\\
               I.\ Antoniadis, C.\ Bachas, and C.\ Kounnas,
	       \NPB{289}{1987}{87};\\
	       I.\ Antoniadis and C.\ Bachas, \NPB{289}{1987}{87}.

\bibitem{fny} \AEF, D.V.\ Nanopoulos and K.\ Yuan,
                                                 \NPB{335}{1990}{347};\\
              G.B.\ Cleaver, \AEF~ and D.V.\ Nanopoulos,
                                                  \PLB{455}{1999}{135}.	

\bibitem{cfr} K. Christodoulides, \AEF~and J. Rizos, \PLB{702}{2011}{81}. 

\bibitem{top} \AEF, \PLB{274}{1992}{47}; \PLB{377}{1996}{43}. 

\bibitem{fmmckm} J. Rizos and K. Tamvakis, \PLB{279}{1992}{281};\\
                 \AEF, \NPB{407}{1993}{57}; \NPB{487}{1997}{55};\\
                 \AEF~and E. Halyo, \PLB{307}{1993}{305}; \NPB{416}{1994}{63}. 

\bibitem{nmasses} I. Antoniadis, J. Rizos and K. Tamvakis, 
                                     \PLB{279}{1992}{281};\\
                  \AEF, \PLB{245}{1990}{435};\\
                  \AEF~and E. Halyo, \PLB{307}{1993}{311};\\
                  C. Coriano and \AEF, \PLB{581}{2004}{99}. 

\bibitem{gcu} \AEF, \PLB{302}{1993}{202}; \\
              K.R. Dienes and \AEF, \PRL{75}{1995}{2646};
                                    \NPB{457}{1996}{409}. 

\bibitem{ps} \AEF, \NPB{428}{1994}{111}; \PLB{520}{2001}{337}.  

\bibitem{squarkdegen} \AEF~and J.C. Pati, \NPB{526}{1998}{21}.

\bibitem{moduli} \AEF, \NPB{728}{2005}{83}.

\bibitem{classification} 
 \AEF, C. Kounnas, S.E.M. Nooij and J. Rizos, \NPB{695}{2004}{41};\\
 \AEF, C. Kounnas and J. Rizos, \PLB{648}{2007}{84}.

\bibitem{psclassi}
 B. Assel, K. Christodoulides, C. Kounnas and J. Rizos, 
 \PLB{683}{2010}{306}; \NPB{844}{2011}{365}. 

\bibitem{fsuclassi}
 \AEF, J. Rizos and H. Sonmez, \NPB{886}{2014}{202}. 


\bibitem{svduality} \AEF, C. Kounnas and J. Rizos, \NPB{774}{2007}{208};
                                                    \NPB{799}{2008}{19};\\
 T. Catelin--Jullien, \AEF, C. Kounnas and J. Rizos, \NPB{812}{2009}{103};\\
 C. Angelantonj, \AEF~ and M. Tsulaia, \JHP{1007}{2010}{004}. 

\bibitem{florakis}
 \AEF, I. Florakis, T. Mohaupt and M. Tsulaia, \NPB{848}{2011}{332}.

\bibitem{ibanezuranga} 
                  L. Ibanez and A. Uranga, {\it String theory and particle
                  physics: an introduction to string phenomenology}, 
                  Cambridge University Press, 2012. 

\bibitem{zpbminusl} \AEF~and D.V. Nanopoulos, \MODA{6}{1991}{61}. 

\bibitem{pati} J.C. Pati, \PLB{388}{1996}{532}. 

\bibitem{slm} \AEF, \PLB{278}{1992}{131}; \NPB{387}{1992}{239}.

\bibitem{plboxone} \AEF, \PLB{499}{2001}{147}.

\bibitem{cfg} C. Coriano, \AEF~ and M. Guzzi, \EJP{53}{2008}{421}; 
                                             \PRD{78}{2008}{015012}. 

\bibitem{chang} D. Chang and A. Kumar, \PRD{38}{1988}{1893}; 
                                       \PRD{38}{1988}{3734}. 

\bibitem{nahe} \AEF~and D.V.\ Nanopoulos, \PRD{48}{1993}{3288}.

\bibitem{atlascms} G.~Aad {\it et al.} [ATLAS Collaboration], 
                                                  arXiv:1506.00962;\\
V.~Khachatryan {\it et al.} [CMS Collaboration], \JHP{1408}{2014}{173};\\
V.~Khachatryan~{\it et al.}~[CMS Collaboration], \JHP{1408}{2014}{174}. 

\bibitem{guzzi} \AEF~and M. Guzzi, arXiv:1507:07406;\\
                T. li, J.A. Maxin, V.E. Mayes and D.V. Nanopoulos, 
                 arXiv:1509.06821.

\bibitem{lrs} G.B. Cleaver, \AEF~and C. Savage, \PRD{63}{2001}{066001};\\
  G.B. Cleaver, D.J. Clements and \AEF, \PRD{65}{2002}{106003};

\bibitem{viraf}\AEF~and V. Mehta, \PRD{84}{2011}{086006}; 
                                    \PRD{88}{2013}{025006}.


\bibitem{fsu5} I.\ Antoniadis, J.\ Ellis, J.\ Hagelin and D.V.\ Nanopoulos
                \PLB{231}{1989}{65};\\
%
\bibitem{alr} 
I.\ Antoniadis, G.K.\ Leontaris and J.\ Rizos,
				\PLB{245}{1990}{161};\\
		G.K.\ Leontaris and J.\ Rizos,
				\NPB{554}{1999}{3}.

\bibitem{cleaverau1} G.B. Cleaver and \AEF, \IJMP{14}{1999}{2335}.


\bibitem{su62}
  L.~Bernard, \AEF, I.~Glasser, J.~Rizos and H.~Sonmez, \NPB{868}{2013}{1}.

\bibitem{afp} P. Athanasopoulos, \AEF~and V. Mehta, \PRD{89}{2014}{105023}.

\bibitem{tduality} A. Giveon, M. Porrati and E. Rabinovici, \PRT{244}{1994}{77}.

\bibitem{mirror} B.R. Greene and M.R. Plesser, \NPB{338}{1990}{15}. 

\bibitem{gepner} D. Gepner, \PLB{199}{1987}{380}; \NPB{296}{1988}{757}.

\bibitem{panos} P. Athanasopoulos, \AEF~ and D. Gepner, \PLB{735}{2014}{357}.

\bibitem{typeIIbclass} A. Gregori, C. Kounnas and J. Rizos, 
                                \NPB{549}{1999}{16}. 

\bibitem{zpsmodel} \AEF~and J. Rizos, \NPB{895}{2015}{233}. 

\bibitem{ssr} S. Chang, C. Coriano and \AEF, \NPB{477}{1996}{65}. 

\bibitem{others} B. Greene, K. Kirklin, P. Miron and G.G. Ross, 
                                           \NPB{278}{1986}{667}; \\
                 L. Ibanez, J.E. Kim, P. Nilles, F. Quevedo, 
                                           \PLB{191}{1987}{282};\\
                D. Gepner, hep-th/9301089;\\
                T. Kobayashi, S. Raby and R. Zhang, \NPB{704}{2005}{3}; \\
                O. Lebedev \etal, \PLB{645}{2007}{88};\\
                M. Blaszczyk \etal, \PLB{683}{2010}{340}; \\
                B. Gato--Rivera, A. Schellekens, \NPB{828}{2010}{375}. 



\end{thebibliography}
\end{document}